# The Highest Energy Emission Detected by EGRET from Blazars

Brenda L. Dingus[1] & David L. Bertsch[2]

(1) Physics Department, University of Wisconsin, Madison, WI 53711
dingus@physics.wisc.edu
(2) NASA Goddard Space Flight Center, Greeenbelt, MD 20771

**Abstract.** Published EGRET spectra from blazars extend only to 10 GeV, yet EGRET has detected approximately 2000 γ-rays above 10 GeV of which about half are at high Galactic latitude. We report a search of these high-energy γ-rays for associations with the EGRET and TeV detected blazars. Because the point spread function of EGRET improves with energy, only ~2 γ-rays are expected to be positionally coincident with the 80 blazars searched, yet 23 γ-rays were observed. This collection of > 10 GeV sources should be of particular interest due to the improved sensitivity and lower energy thresholds of ground-based TeV observatories. One of the blazars, RGB0509+056, has the highest energy γ-rays detected by EGRET from any blazar with 2 > 40 GeV, and is a BL Lac type blazar with unknown redshift.

## INTRODUCTION

EGRET catalogs have been produced in different energy ranges, > 100 MeV [1] and > 1 GeV [2]. However, EGRET also detected approximately 2000 γ-rays above 10 GeV. At these high energies the number of γ-rays detected is small due to both the source flux and EGRET's effective area decreasing with energy. The loss of detection efficiency is caused by higher energy γ-rays producing showers in the calorimeter with particles that propagate backwards and hit the anticoincidence dome, and thus the trigger is self-vetoed. The energy resolution is ~20-50% at these energies because of shower leakage out of the calorimeter. However, the point spread function (psf) dramatically improves with energy going from ~6$^o$ at 100 MeV (68% containment) to 0.3$^o$ at 10 GeV. The preflight calibration of the psf is shown in Figure 1 and the locations of the 35 >10 GeV γ-rays near the EGRET-detected pulsars are plotted in Figure 2. More information about the Galactic sources is given in an accompanying paper [3]. The subject of this paper is a seach for positional coincidences between the ~1000 γ-rays more than 10 degrees from the Galactic plane and the sample of 77 active galactic nuclei (AGN) detected by EGRET and 3 additional AGN (1ES2344+514, 1ES1959+650, 1ES1426+428) detected only at TeV energies.

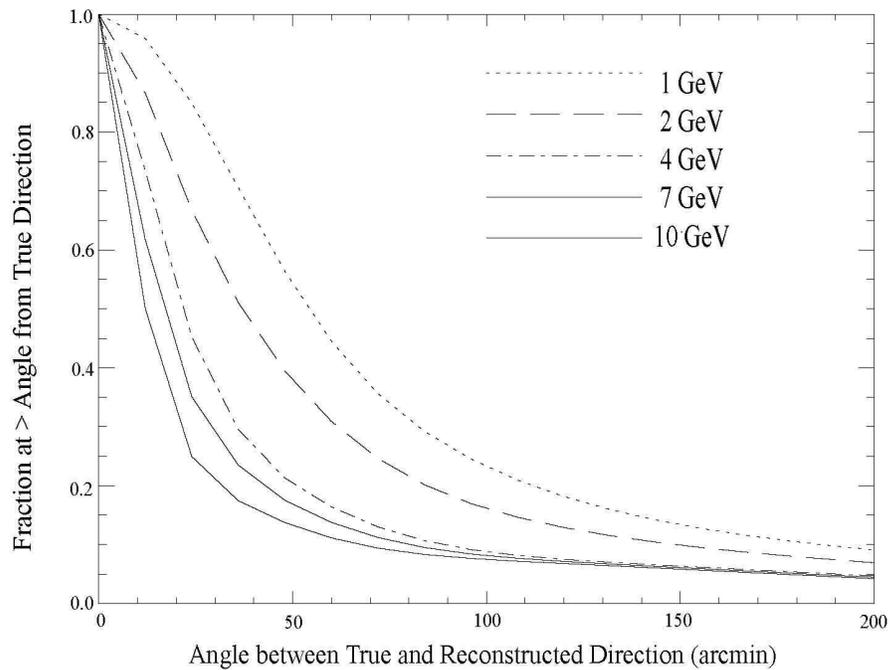

**FIGURE 1.** The EGRET point spread function for γ-rays of different energies as determined by the pre-flight calibration.

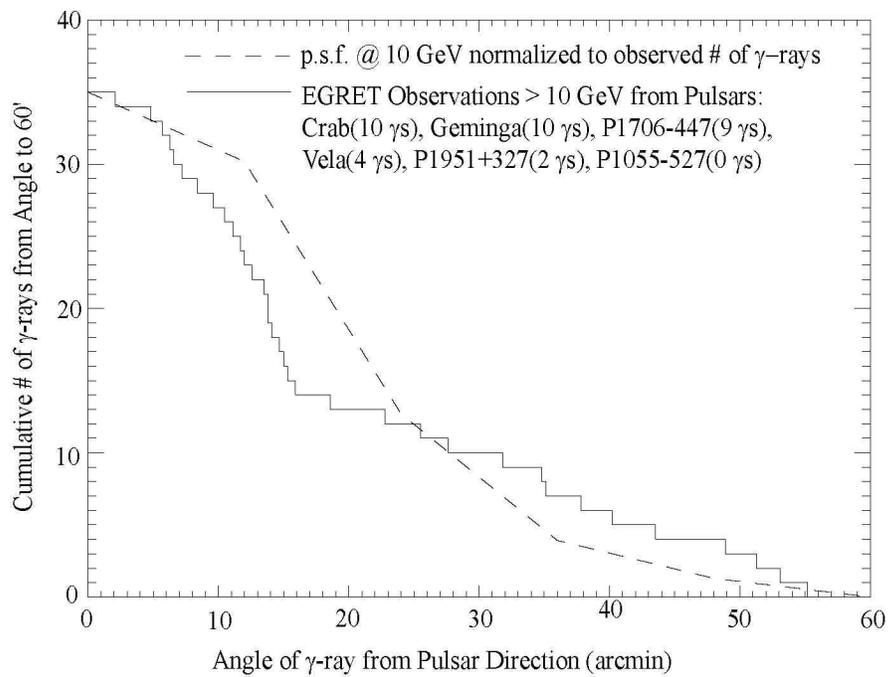

**FIGURE 2.** The direction of the 35 γ-rays > 10 GeV within 1° of the EGRET-detected pulsars. The dashed line corresponds to the 10 GeV point spread function of Figure 1, but normalized to the number of γ-rays detected.

# SEARCH TECHNIQUE & RESULTS

The improved angular resolution and the small number of γ-rays and sources allow a simple search technique. All γ-rays within 30 arcmin of any of the 80 active galactic nuclei known to emit high-energy γ-rays are selected. The search region of 30 arcmin radius is chosen to include a large fraction of the point spread function as shown in Figures 1 and 2. The number of positional coincidences occurring by chance can be estimated by multiplying the solid angle of the search region, 80 x π (30 arcmin$^2$), by the average density of γ-rays more than 10 degrees from the Galactic plane, ~ 1000 γ / 4π(1-sin10$^o$) sr, which gives ~2 γ-rays expected due to random coincidences. A more detailed calculation using the EGRET exposure as a function of Galactic latitude and longitude plus a model of the Galactic diffuse γ-rays [5] and the isotropic diffuse γ-ray flux [6] normalized to the observed number of γ-rays, yielded a similar number of 2.2 γ-rays expected by chance in the 80 regions surrounding these AGN. Since 23 γ-rays were found to be positionally coincident with the 80 AGN, the probability is <10$^{-16}$ that these γ-rays are chance coincidences.

The 23 selected γ-rays are listed in Table 1. The characteristics of the sources and the γ-rays are listed as well as 3 probabilities. The probability #1 is the chance that a single γ-ray > 10 GeV will be within 30 arcmin of that AGN by random chance. This calculated number depends on the source location and is determined from the EGRET exposure and diffuse models mentioned in the previous paragraph. The probability #2 is not a true a priori chance probability, but is an attempt to identify which sources are most likely to emit higher energy γ-rays. This number is calculated by multiplying the probability #1 by the number of γ-rays with the energy of γ-ray detected or higher divided by the number of γ-rays > 10 GeV, as well as multiplying by the square of the ratio of the distance between the γ-ray detected and 30 arcmin. Thus the higher the energy and the closer the γ-ray is to the source location, the lower this second probability. The probability #3 is given only when more than one γ-ray is associated with a single source. This probability #3 is the chance of getting the combination of the #2 probabilities, $\alpha_i$ for i=1 to N, the number of γ-rays detected from that source, and is determined assuming –2 ln Π $\alpha_i$ is distributed as a $\chi^2$(2N). The list is sorted such that the 5 sources with more than one γ-ray are listed first, and the remaining sources are ordered by increasing probability #2.

Several of these sources are not extensively studied at lower energies, and some do not have known redshifts. For example, the first source in the table RGB0506+059 [7] has 2 γ-rays above 40 GeV, but is not detected in the 3$^{rd}$ EGRET catalog [1], because it does not emit a significant flux in the energy range above 100 MeV. It is a 4.5 σ detection in the >1GeV catalog [2] with only 23±7 γ-rays detected from the source, and a flux >1 GeV of (1.4 ± 0.4) x 10$^{-8}$ γ/cm$^2$ sec. The flux above 10 GeV flux is very uncertain due to the statistical error with only 2 γ-rays detected, and due to the systematic errors in calculating the self-veto by the anticoincidence shield.

**TABLE 1. All γ-rays >10 GeV within 0.5 degrees of EGRET or TeV detected blazars.**

| AGN Properties | | | | Gamma-Ray Properties | | | Chance Probabilities See Notes | | |
|---|---|---|---|---|---|---|---|---|---|
| Name | Gal. Long. | Gal. Lat. | z | Date | Energy (GeV) | Arcmin From AGN | #1 | #2 | #3 |
| 0509+056 | 195.40 | -19.64 | ? | Feb-94 | 45.1 | 3.0 | 6.9E-02 | 4.1E-05 | 5.0E-08 |
|  |  |  |  | Aug-94 | 44.9 | 3.6 | 6.9E-02 | 5.9E-05 |  |
| 2155-304 | 17.73 | -52.25 | 0.116 | Jan-98 | 11.0 | 7.8 | 6.2E-03 | 3.5E-04 | 2.8E-08 |
|  |  |  |  | Feb-93 | 11.1 | 12.6 | 6.2E-03 | 9.2E-04 |  |
|  |  |  |  | May-96 | 11.2 | 7.2 | 6.2E-03 | 3.0E-04 |  |
| 0430+285 | 170.52 | -12.60 | ? | Aug-95 | 15.7 | 6.0 | 1.7E-01 | 3.1E-03 | 4.5E-05 |
|  |  |  |  | Aug-96 | 14.1 | 6.6 | 1.7E-01 | 4.6E-03 |  |
|  |  |  |  | Feb-97 | 29.2 | 30.0 | 1.7E-01 | 2.5E-02 |  |
| 1406-076 | 333.88 | 50.28 | 1.494 | Jan-93 | 11.2 | 22.8 | 1.1E-02 | 2.7E-03 | 1.7E-04 |
|  |  |  |  | Dec-94 | 10.1 | 15.0 | 1.1E-02 | 5.2E-03 |  |
| 1622-253 | 352.14 | 16.32 | 0.786 | Dec-91 | 14.2 | 14.4 | 1.6E-01 | 2.0E-02 | 2.7E-03 |
|  |  |  |  | Sep-93 | 19.4 | 16.2 | 1.6E-01 | 1.5E-02 |  |
| 1219+285 | 201.74 | 83.29 | 0.102 | Apr-93 | 27.3 | 7.2 | 7.6E-03 | 7.5E-05 |  |
| 0208-512 | 276.10 | -61.78 | 1.003 | Mar-96 | 10.7 | 11.4 | 9.7E-03 | 1.2E-03 |  |
| 1101+384 | 179.83 | 65.03 | 0.031 | Sep-92 | 14.2 | 19.8 | 6.7E-03 | 1.6E-03 |  |
| 1730-130 | 12.03 | 10.81 | 0.902 | Apr-94 | 23.7 | 7.2 | 1.6E-01 | 2.0E-03 |  |
| 1759-396 | 352.45 | -8.43 | ? | Jun-94 | 29.6 | 10.8 | 2.0E-01 | 2.0E-03 |  |
| 1611+343 | 55.15 | 46.38 | 1.401 | Nov-92 | 10.9 | 19.8 | 6.2E-03 | 2.3E-03 |  |
| 1622-297 | 348.82 | 13.32 | 0.815 | Apr-94 | 13.2 | 6.0 | 1.1E-01 | 2.6E-03 |  |
| 1633+382 | 61.09 | 42.34 | 1.814 | Sep-91 | 11.0 | 20.4 | 7.1E-03 | 2.8E-03 |  |
| 0440-003 | 197.20 | -28.46 | 0.844 | Aug-94 | 15.4 | 24.0 | 1.9E-02 | 5.6E-03 |  |
| 0219+428 | 140.14 | -16.77 | 0.444 | Oct-98 | 12.2 | 26.4 | 2.1E-02 | 1.1E-02 |  |
| 1322-428 | 309.52 | 19.42 | 0.0018 | Apr-93 | 10.0 | 26.4 | 3.7E-02 | 2.9E-02 |  |

Notes: #1 is chance probability that a diffuse γ-ray > 10 GeV is within 0.5 degrees from that AGN.
#2 is chance probability that a diffuse γ-ray > energy detected is within the distance detected from that AGN.
#3 is combined probability of multiple γ-rays with chance probability #2. (See text for the formula.)

However, a comparison with the Crab observation implies the flux is of order one-tenth the Crab flux at these energies. This blazar is observed optically with a magnitude of 15.3. The radio flux is variable with the flux density as high as 1 Jy at 4.85 GHz and the radio spectrum is flat. ROSAT detected the source with a flux of $(1.74 \pm 0.47) \times 10^{-15}$ W/m$^2$ in the all sky survey [8].

## DISCUSSION

The AGN listed in Table 1 are a better indicator of the highest energy γ-ray emitters than the extrapolations of the spectra from the 3$^{rd}$ EGRET catalog [1]. These spectral fits have large uncertainties and there may be spectral changes before these energies. TeV sources have proven to be weak in the EGRET energy range implying a steepening of the spectral slope at energies above EGRET's detections. In fact, many bright >100 MeV EGRET sources, such as 3C279 or PKS0528+134, are not in this table. Some of the sources in the table have been reported as TeV sources, such as 1101+384 (Mrk 421), PKS2155-304, and 0219+428 (3C66A) [4]. This list points to several AGN that should be detectable by new ground-based observatories which have increased sensitivity and lower energy thresholds of ~100 GeV. GLAST will also have a much larger area and field of view, and will be able to better measure the flux and spectral features such as spectral breaks due to the source or due to pair production on the extragalactic background light.

## ACKNOWLEDGMENTS

The author acknowledges support from NASA, NSF and Research Corporation. This research has made use of the NASA / IPAC Extragalactic Database (NED) which is operated by the Jet Propulsion Laboratory, California Institute of Technology, under contract with NASA.